\documentclass[aps,prl,%
    amssymb,amsmath,%
    superscriptaddress,%
    noshowpacs,noshowkeys,twoside,twocolumn,letterpaper,floatfix,%
    ]{revtex4}

\usepackage{graphicx}

\newcommand*{\ket}[1]{|{#1}\rangle}
\newcommand*{\bra}[1]{\langle{#1}|}

\newcommand*{\textgap}{{\text{gap}}}
\newcommand*{\textknight}{{\text{K}}}

\begin{document}
\title{Qubit protection in nuclear-spin quantum dot memories}
\author{Z. Kurucz}
\affiliation{Fachbereich Physik, University of Kaiserslautern, D-67663 Kaiserslautern, Germany}
\affiliation{Research Institute for Solid State Physics and Optics, 
  H.A.S., H-1525 Budapest, Hungary}
\author{M. W. S{\o}rensen}
\affiliation{Niels Bohr Institute, University of Copenhagen, DK-2100 Copenhagen, Denmark}
\author{J. M. Taylor}
\affiliation{Department of Physics, Massachusetts Institute of Technology, Cambridge, Massachusetts 02139, USA}
\affiliation{Department of Physics, Harvard University, Cambridge, Massachusetts 02138, USA}
\author{M. D. Lukin}
\affiliation{Department of Physics, Harvard University, Cambridge, Massachusetts 02138, USA}
\author{M. Fleischhauer}
\affiliation{Fachbereich Physik, University of Kaiserslautern, D-67663 Kaiserslautern, Germany}
\begin{abstract}
  We present a mechanism to protect quantum information stored in an
  ensemble of nuclear spins in a semiconductor quantum dot.  When the
  dot is charged the nuclei interact with the spin of the excess
  electron through the hyperfine coupling. If this coupling is made
  off-resonant it leads to an energy gap between the collective
  storage states and all other states. We show that the energy gap
  protects the quantum memory from local spin-flip and spin-dephasing
  noise.  Effects of non-perfect initial spin polarization and
  inhomogeneous hyperfine coupling are discussed.
\end{abstract}
\maketitle

An essential ingredient for quantum computation and long-distance
quantum communication is a reliable quantum memory.  Nuclear spins in
semiconductor nanostructures are excellent candidates for this task.
With a magneton 3 orders of magnitude weaker than electron spins, they
are largely decoupled from their environment, and the hyperfine
interaction with electron spins allows one to access ensembles of
nuclear spins in a controlled way \citep{prl91e017402, prb75e155324,
  prl90e206803, prl91e246802, sci316p1312, sci320p1326, prb73e245318,
  prl96e167403, pra74e032316}.  In particular, the quantum state of an
electron spin can be mapped onto the nuclear spins, giving rise to a
long-term memory \citep{prl90e206803, prl91e246802, sci316p1312,
  sci320p1326}.  Nevertheless, memory lifetimes are limited, e.g., by
dipole-dipole interactions among the nuclei.  In this Letter we
demonstrate that the presence of the electron spin in the quantum dot
substantially reduces the decoherence of this collective memory
associated with surrounding nuclear spins.  The virtual transitions
between electronic and nuclear states can be used to produce an energy
shift proportional to the number of excitations in the storage
spin-wave mode. This isolates the storage states energetically and
protects them against nuclear spin flips and spin diffusion.

\begin{figure}
  \includegraphics{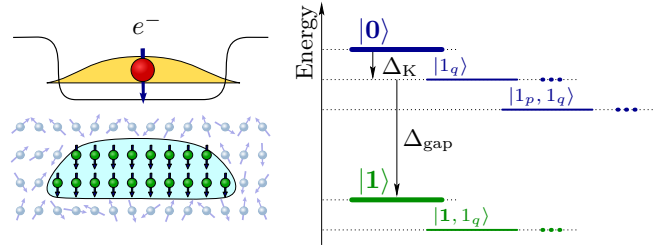}
  \caption{
    (Color online) \textit{Left:} Charged quantum dot with a single,
    polarized excess electron.  \textit{Right:} Spectrum of the
    effective nuclear Hamiltonian in the presence of a polarized
    electron.  Off-resonant hyperfine coupling results in a gap
    $\Delta_\textgap$ between the storage state $\ket{\mathbf 1}$ and
    the non-storage states $\ket{1_q}$.  $\Delta_\textknight$ denotes
    the Zeeman shift due to the effective magnetic field associated
    with the electron spin (Knight shift).
  }
  \label{fig:q-dot-system}
\end{figure}

Consider a quantum dot charged with a single excess electron as
indicated in Fig.~\ref{fig:q-dot-system}.  The electron spin $\hat
{\mathbf S}$ is coupled to the ensemble of underlying nuclear spins
$\hat {\mathbf I}^j$ by the Fermi contact interaction,
\begin{gather}
  \label{eq:Hint-hyperfine}
  \hat H_{\text{hf}}
  = \mathcal A \sum_j^N 
  \varrho_j \Big[ \hat I^{j}_z \hat S_z 
  + \tfrac12 \Big( \hat I^{j}_+ \hat S_-  
  + \hat I^{j}_- \hat S_+ \Big)\Big],
\end{gather}
where $\mathcal A$ is the average hyperfine interaction constant,
$\mathcal A \approx 90\,\mathrm{\mu eV}$ for GaAs, and $\varrho_j$ is
proportional to the electron density at the position of the $j$th
nucleus, $\sum_j \varrho_j =1$.  For convenience, we introduce the
collective operators $\hat{\mathbf A} \equiv \sum_j \varrho_j
\hat{\mathbf I}^j$.  The first term in Eq.~\eqref {eq:Hint-hyperfine}
provides an effective magnetic field $B_z^{\text{OH}} = \mathcal A
\langle \hat A_z \rangle / g^* \mu_B$ for the electron, known as the
Overhauser field.  The same also produces an energy shift for each
nuclei, the so-called Knight shift.  The flip-flop terms in Eq.~\eqref
{eq:Hint-hyperfine}, $\hat H_{\text{JC}} = \frac{\mathcal A}2 (\hat
A_+ \hat S_- + \hat A_- \hat S_+)$, can be used to polarize the
nuclear spins~\citep{prl91e017402, prb75e155324}, and to map the
electron's spin state into a collective spin mode of the nuclei
\citep{prl90e206803, prl91e246802}.  As will be shown here, the same
can be used to provide a protective energy gap.

\paragraph{Fully polarized nuclei.}%
We start by reconsidering the storage of a qubit in a collective
nuclear state~\citep{prl90e206803}.  In the simplest case when all the
nuclear spins are initially polarized in the $-z$ direction (zero
temperature limit), the $\ket\downarrow_e$ and $\ket\uparrow_e$ spin
states of the electron are mapped onto the nuclear spin states
\begin{align}
  \label{eq:ket0}
  \ket{\mathbf 0} &\equiv \ket{-I, -I, \ldots, -I},\\
  \label{eq:ket1}
  \ket{\mathbf 1} &\equiv
     \tfrac{\mathcal A}\Omega 
    \hat A_+ \ket{\mathbf 0}
    \propto \sum_j \varrho_j 
    \ket{-I, \ldots, (-I+1)_j, \ldots, -I},
\end{align}
respectively.  $\hat H_{\text{JC}}$ couples the state $\ket{\mathbf 0}
\ket\uparrow_e$ to $\ket{\mathbf 1} \ket\downarrow_e$ with an angular
frequency $\Omega = \mathcal A \big({\sum_j \varrho_j^22I}\big)
{}^{1/2}$.  The detuning between these two states, $\delta =
\delta^{\text{el}} + \delta^{\text{OH}}$, comes from the electron's
intrinsic energy splitting $\delta^{\text{el}}$ due to, e.g., an
external magnetic field, and from the Overhauser field,
$\delta^{\text{OH}} = -\mathcal AI$.  Coherent flip-flops between the
electron and nuclear spins can be brought into resonance ($\delta \ll
\Omega$) through $\delta^{\text{el}}$, e.g., applying a spin-state
dependent Stark laser pulse~\citep{prb63e085303}.  Then
$\ket{\mathbf0} (\alpha \ket\downarrow_e + \beta \ket\uparrow_e)$ can
be rotated to $(\alpha \ket{\mathbf0} + \beta \ket{\mathbf1})
\ket\downarrow_e$, and the quantum information can be transferred from
the electron to the nuclear spin ensemble and back~\citep
{prl90e206803, prl91e246802}.

Assume that, after the qubit has been written into the nuclei, the
polarized electron is not removed from the dot but the hyperfine
flip-flops are tuned off-resonant ($\delta\gg\Omega$).  Now real
transitions can no longer take place between $\ket{\mathbf 1}
\ket\downarrow_e$ and $\ket{\mathbf 0} \ket\uparrow_e$.  However, the
residual virtual transitions repel the two states from each other, in
analogy to the dynamic Stark effect.  As a result, after eliminating
the electron, the energy of state $\ket{\mathbf 1}$ gets shifted by
$\Delta_\textgap = -\Omega^2/4\delta$.  The other, orthogonal states
also having exactly one spin flipped (denoted by $\ket{1_q}$ in
Fig.~\ref {fig:q-dot-system}) are ``subradiant'', i.e., are not
coupled via $\hat H_{\text{JC}}$ to the electron.  Therefore, they are
unaffected by the shift.  This is the origin of the energy gap.

To understand the protection scheme, let us introduce \emph{nuclear
  spin waves}.  As long as the nuclei remain highly polarized, one can
introduce bosonic operators through the Holstein-Primakoff
transformation: $\hat a_j \approx \hat I_-^j/ \sqrt{2I}$, $\hat
a_j^\dagger \approx \hat I_+^j/\sqrt{2I}$, and $\hat a_j^\dagger \hat
a_j = \hat I_z^j + I$.  This allows us to define the bosonic spin
waves
\begin{equation}
  \label{eq:spin-modes}
  \hat \Phi_q\equiv \sum_j \eta_{qj} \hat a_j,\qquad
  \hat \Phi_q^\dagger  \equiv \sum_j \eta_{qj}^* \hat a_j^\dagger,
\end{equation}
where the unitary matrix $\eta_{qj}$ describes the mode functions.  We
identify the storage mode $q=0$ as the one given by $\eta_{0j} =
\sqrt{2I} \frac {\mathcal{A}} \Omega \varrho_j$, and write
$\ket{\mathbf 1} = \hat \Phi_0^\dagger \ket{\mathbf 0}$.  This is the
mode which is directly coupled to the electron spin.  In fact, $\hat
H_{\text{JC}} \approx \frac\Omega2 \big( \hat\Phi_0^\dag \hat S_- +
\hat\Phi_0 \hat S_+ \big)$ is a Jaynes-Cummings coupling in the
bosonic approximation.  After eliminating the electron, $\hat
H_{\text{JC}}$ reduces to $\hat H_\textgap = -\frac{\mathcal
  A^2}{4\delta} \hat A_+ \hat A_- \approx \Delta_\textgap \hat
\Phi_0^\dagger\hat \Phi_0$.
As shown in Fig.~\ref{fig:q-dot-system}, $\hat H_\textgap$ lifts the
degeneracy between states of different number of storage-mode
excitations. This is the key feature of our protection scheme: any
decoherence process that is associated with a transition from the
storage mode $\hat \Phi_0$ to any other mode $\hat \Phi_q$ now has to
bridge an energy difference.  If this gap is larger than the spectral
width of the noise, the effect of the noise is substantially reduced.

A more detailed analysis shows that the off-resonant interaction with
the electron spin---which itself is coupled, e.g., to phonons---leads
in general also to an additional decoherence mechanism for the nuclear
spins.  If the corresponding electron spin dephasing rate $\gamma$ is
small compared to the electron's precession frequency $\delta$, the
decay rate for the storage mode is reduced by the low probability of
exciting the electron spin state: $\gamma \Omega^2 / \delta^2 \ll
\gamma$.

In addition to the gap, the electron is also responsible for the
Knight shift $\hat H_\textknight = \mathcal A \hat A_z \langle \hat
S_z \rangle$.  The difference of the Knight shifts for the
$\ket{\mathbf0}$ and $\ket{\mathbf1}$ states, $\Delta_\textknight =
-\frac{\mathcal A}2 \sum_j \varrho_j^3 \big/ \sum_j \varrho_j^2$, is
typically much less than $\Delta_\textgap$.  When the hyperfine
coupling is \emph{inhomogeneous}, however, $\ket{\mathbf1}$ fails to
be eigenstate of the Knight shift Hamiltonian: $\hat H_\textknight
\ket{\mathbf1} = (-\frac12 \delta^{\text{OH}} + \Delta_\textknight)
\ket{\mathbf1} + \zeta \ket{1^\perp}$, where the state $\ket{1^\perp}$
is orthonormal to $\ket{\mathbf1}$ and the coupling parameter $\zeta
^2= \frac{\mathcal A^2}4 \sum_j \varrho_j^4 \big/ \sum_j \varrho_j^2
-\Delta_\textknight^2$ characterizes the inhomogeneities.  As a
consequence, the storage mode is only an approximate eigenmode, and it
gradually mixes with non-storage modes as time passes.  This causes
loss of the stored qubit.  $\ket{1^\perp}$ is, however, off-resonant
due to the energy gap, and our simulations show that the corresponding
probability of finding the system in state $\ket{1^\perp}$ is bounded
by $4\zeta^2 / \Delta_\textgap^2$, so the detrimental effect of the
inhomogeneous Knight shift is suppressed by the energy gap.  In
addition, since the admixture of $\ket{1^\perp}$ is a coherent
process, it can be cancelled by refocusing (echo) methods.

\begin{figure}[b]
  \centerline{\includegraphics[width=26em]{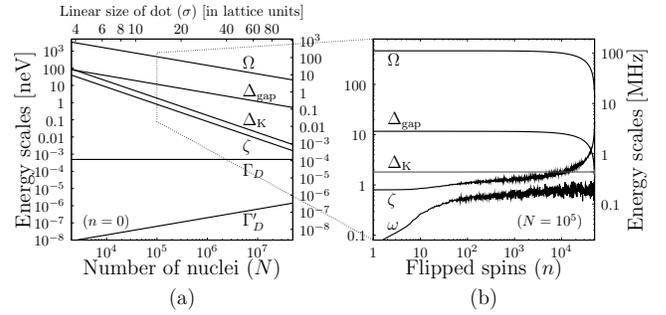}}
  \caption{
    Hyperfine Rabi frequency ($\Omega$), protective energy gap
    ($\Delta_\textgap$), Knight shift difference between the logical
    states ($\Delta_\textknight$), symmetry breaking couplings due to
    inhomogeneities ($\zeta$ and $\omega$), qubit decoherence rate due
    to dipolar spin diffusion without ($\Gamma_D$) and with
    ($\Gamma_D'$) protection.  (a) The fully polarized (zero
    temperature) case is displayed as function of the number of
    spin-$\frac12$ nuclei ($N$) taking part in the storage, i.e.,
    located within $3\sigma$ of the oblate Gaussian electron
    distribution with in-plane variance~$\sigma$.  (b) Estimated
    energies in dark states $\ket{\mathcal D_{n,\beta}}$ with $n$
    spins flipped from the fully polarized state for $N=10^5$.  Energy
    units are chosen to match GaAs.
  }
  \label{fig:params}
\end{figure}

A large gap can be achieved by bringing the hyperfine interaction
close to resonance.  For example, a non-zero external magnetic field
or laser induced AC Stark shifts \citep{prb63e085303} can partially cancel
the Overhauser field, such that $\delta \ll \delta_{\text{el}} \approx
-\delta_{\text{OH}} = \mathcal AI$.  (Of course, $\delta$ should be
kept sufficiently large so that the hyperfine coupling remains
off-resonant).  The requirement of separation of time scales implies
$\zeta \ll |\Delta_\textgap| \ll \Omega \ll |\delta|$,
i.e., $\delta\gtrsim10\Omega$.  To estimate the orders of
magnitude of the different energies, we take an oblate Gaussian
electron density of ratio $(1,1,1/3)$, and we consider spin-$\frac12$
nuclei.  Then it is easy to see that $\Delta_\textknight$ and~$\zeta$
are inversely proportional to the number of nuclei~$N$, whereas
$\Omega, \Delta_\textgap \propto N^{-1/2}$ only (Fig.~\ref
{fig:params}a).

To analyze the \emph{decoherence suppression}, we first consider a
simplistic noise model where the nuclear spins are coupled to
fluctuating, classical fields.  The corresponding interaction
Hamiltonian is given by $\hat V = \sum_j {\mathbf B}^j\cdot \hat
{\mathbf I}^j$.  We assume isotropic Gaussian noise with zero mean and
\begin{gather}
  \label{eq:noise-correlator}
  \overline{ B^j_\mu(t) B^k_\nu(t')} = \delta_{\mu \nu}\, \xi_{jk} 
  C e^{-\Gamma |t-t'|}
\end{gather}
for $\mu,\nu = x,y,z$, where $\xi_{jk}$ specifies the spatial
correlations of the noise acting on different nuclei.  For simplicity,
the noise spectrum is assumed to be Lorentzian with a width $\Gamma$,
although similar results hold for other spectra with a high-frequency
cut-off.

Let us first discuss the \emph{dephasing part}, $\hat V_z = \sum_j
B_z^j \hat I^j_z$, of the noise.  Using the bosonic spin-wave
operators introduced in Eq.~\eqref{eq:spin-modes} we can express $\hat
V_z$ as
\begin{align}
  \label{eq:V_z}
  \hat V_z = \sum_j B_z^j \hat a_j^\dag \hat a_j
  = \sum_{pq} 
  \bigg( \sum_j B^j_z \eta_{pj}^* \eta_{qj} \bigg)
  \hat\Phi_p^\dag \hat\Phi_q.
\end{align}
Dephasing of individual nuclear spins thus means transfer of
excitations between different spin-wave modes.  Especially, it leads
to both real and virtual transitions from $\ket{\mathbf1}$ to a
non-storage state $\ket{1_q}$ (with $q\ne0$).  As the latter state is
``subradiant'' and, thus, equivalent to $\ket{\mathbf0}$ when the
memory is read out, this process essentially results in damping (for
real transitions) and dephasing (for virtual transitions) of the
stored logical qubit~\citep{pra72e022327&Mewes-dissertation}.  This
can be seen by formally eliminating the classical fields and all
non-storage mode in Markov approximation and deriving a master
equation for the storage mode.  For that, we assume the zero
temperature limit with all non-storage modes $\hat \Phi_{q\ne 0}$ in
the vacuum state.  This results in
\begin{gather}
  \label{eq:noise-master-reduced}
  \frac d{dt} \hat\rho
  = i \big[\hat\rho, E_{z} \hat\Phi_0^\dag \hat\Phi_0 \big]
  + \mathcal L_{z} (\hat\rho),
\end{gather}
with energy shift $ E_{z} = (1-\Xi){C \Delta_\textgap} / {(\Gamma^2 +
  \Delta_\textgap^2)}$ and
\begin{multline}
  \mathcal L_{z} (\hat\rho) 
  = {\gamma_1} \big( 
  2\hat\Phi_0 \hat\rho \hat\Phi_0^\dag
  -\hat\Phi_0^\dag \hat\Phi_0 \hat\rho
  -\hat\rho \hat\Phi_0^\dag \hat\Phi_0 \big)  
  \\
  + {\gamma_2} \big( 
  2\hat\Phi_0^\dag \hat\Phi_0 \hat\rho \hat\Phi_0^\dag \hat\Phi_0
  -\hat\Phi_0^\dag \hat\Phi_0 \hat\Phi_0^\dag \hat\Phi_0 \hat\rho 
  -\hat\rho \hat\Phi_0^\dag \hat\Phi_0 \hat\Phi_0^\dag \hat\Phi_0
  \big).
\end{multline}
Here, $\gamma_1$ is the damping rate of the stored qubit while
$\gamma_2$ describes its dephasing.  The two rates are given by
\begin{equation}
  \gamma_1 = \frac {C\Gamma}{\Gamma^2 + \Delta_\textgap^2} (1-\Xi),
  \qquad
  \gamma_2 = \frac C\Gamma \Xi,
\label{eq:gamma}
\end{equation}
where we have introduced the dimensionless parameter $\Xi \equiv
\sum_{jk} \xi_{jk} \varrho_j^2 \varrho_k^2 \big/ \big( \sum_l\varrho_l^2
\big) {}^2$ containing the spatial part of the noise correlator.

\begin{figure}
  \includegraphics[width=26em]{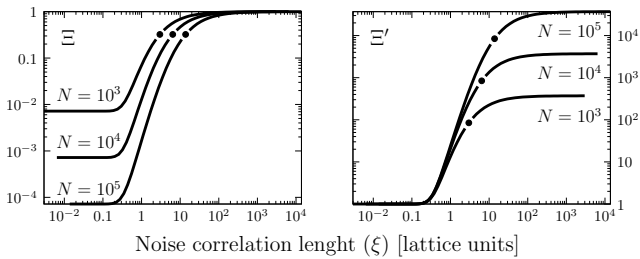}
  \caption{
    The parameters $\Xi$ and $\Xi'$ describing the effects of
    spatial correlations in the classical noise ($\xi_{jk} =
    e^{-r_{jk}/\xi}$) for different number of nuclei.  The same family
    of Gaussian electron densities was used as in
    Fig.~\protect\ref{fig:params}.  The bullets on the curves denote
    the linear size of the dot given by the variance~$\sigma$.
  }
  \label{fig:FF}
\end{figure}

When the correlation length of the classical noise is smaller than the
distance between the nuclei (local uncorrelated noise, $\xi_{jk}\sim
\delta_{jk}$), $\Xi$ scales inversely with the number of nuclei
(Fig.~\ref{fig:FF}).  In this case, the dephasing rate $\gamma_2$
vanishes as $1/N$, which is an effect of the collective nature of the
storage states \cite{pra72e022327&Mewes-dissertation}. The storage of
a qubit corresponds to an encoding of the logical state in a large,
delocalized ensemble of $N$ physical spins. As the decoherence has
strongly local character, there is only a very small effect on the
dephasing of the qubit.  Secondly, the loss of
the stored qubit is due to transitions among states with different
number of excitations in the storage mode.  These transitions are
strongly suppressed and the damping rate $\gamma_1$ is decreased if
$\Delta_\textgap$ is large compared to the width of the noise spectrum
$\Gamma$ (or the corresponding cut-off frequency).  Finally, we note
that the opposite limit of infinite spatial correlation length
($\xi_{jk}=1$) corresponds to a homogeneous random field resulting,
e.g., from a global external source.  In that case, $\Xi \approx 1$
(see Fig.~\ref{fig:FF}) and there is no protection against dephasing.

Following a similar but slightly more involved procedure we can
discuss the \emph{spin-flip part} $\hat V_{xy} = \frac12 \sum_j \big(
B^j_+ \hat I^j_- + B^j_- \hat I^j_+ \big)$ of the noise.  When
deriving a master equation for this case, we need to keep higher order
terms in the Holstein-Primakoff approximation: in the next order $\hat
I^j_- \approx \sqrt{2I} \big( 1 - \lambda \hat a_j^\dag \hat a_j \big)
\hat a_j$ (and similarly for $\hat I^j_+$) with $\lambda =
1-(1-1/2I)^{1/2}$.  Here we have neglected the probability of double
or more excitations on the same site~$j$, which is reasonable in the
high polarization ($T=0$) limit and exact for spin-$\frac12$ nuclei.
Omitting the energy shifts, the Lindbladian describing decoherences
due to spin flips reads, in leading order of~$1/N$,
\begin{multline}
  \mathcal L_{xy} (\hat\rho) 
  =
  (\gamma_3 + \gamma_4) \big(
  2 \hat\Phi_0 \hat\rho \hat\Phi_0^\dag
  - \hat\Phi_0^\dag \hat\Phi_0 \hat\rho
  - \hat\rho \hat\Phi_0^\dag \hat\Phi_0 
  \big)
  \\
  + \gamma_5 \big( 
  2\hat\Phi_0^\dag \hat\Phi_0 \hat\rho \hat\Phi_0^\dag \hat\Phi_0
  -\hat\Phi_0^\dag \hat\Phi_0 \hat\Phi_0^\dag \hat\Phi_0 \hat\rho 
  -\hat\rho \hat\Phi_0^\dag \hat\Phi_0 \hat\Phi_0^\dag \hat\Phi_0
  \big)
  \\
   + \gamma_3 \big( 
  2 \hat\Phi_0^\dag \hat\rho \hat\Phi_0
  - \hat\Phi_0 \hat\Phi_0^\dag \hat\rho 
  - \hat\rho \hat\Phi_0 \hat\Phi_0^\dag
  \big),
\end{multline}
which describes decay with rate $\gamma_4$, dephasing with rate
$\gamma_5$, and additionally thermalization (relaxation to the
identity matrix) with rate $\gamma_3$.  The rates read
\begin{gather}
  \label{eq:gamma345}
  \gamma_3 = \frac {C\Gamma I \Xi'}{\Gamma^2 
    + (\Delta_\textgap + \Delta_\textknight)^2},
  \quad
  \gamma_4 = \frac {2C\Gamma I \lambda^2}{\Gamma^2 
    + (\Delta_\textgap - \Delta_\textknight)^2},
  \nonumber
  \\
  \gamma_5 = \frac {4C\Gamma I \lambda^2}{\Gamma^2 
    + \Delta_\textknight^2}
  \frac{\sum_j \varrho_j^4}{\big(\sum_j \varrho_j^2\big){}^2}.
\end{gather}
In the limit of vanishing spatial correlations of the spin-flip noise,
$\Xi' \equiv \sum_{jk} \xi_{jk} \varrho_j \varrho_k / \sum_l
\varrho_l^2$ tends to $1$ (Fig.~\ref{fig:FF}) and we have protection
against thermalization ($\gamma_3$) because of the separation of
$\ket{\mathbf0}$ and $\ket{\mathbf1}$ by $\Delta_\textgap +
\Delta_\textknight$.  The decay corresponding to $\gamma_4$ is due to
spin-flip induced transitions between $\ket{\mathbf1}$ and $\ket{1_p,
  1_q}$ (the latter containing a total of two excitations but none in
the storage mode), and the energy to bridge is in the order of
$\Delta_\textgap - \Delta_\textknight$ (see Fig.~\ref
{fig:q-dot-system}).  Finally, the last factor in the dephasing rate
$\gamma_5$ scales as $1/N$, indicating that it is the collective
nature of the storage that leads to protection.  Note that the
nonlinearity of the Holstein-Primakoff representation is responsible
for this dephasing: the virtual non-storage excitations are
interacting with the storage mode.

Another potential source of decoherence is \emph{nuclear spin
  diffusion} due to dipole-dipole interaction between nuclear
spins~\citep{ieeetn4p35}.  The energy gap gives protection against
this effect, too.  The dipolar interaction between the pairs of spins
is described in the secular approximation by
\begin{gather}
  \label{eq:dipole}
  \hat H_D = \sum_{j\ne k} B_{jk} 
  \big( \hat I^j_+ \hat I^k_- - 2\hat I^j_z \hat I^k_z\big) 
  \approx 2I \sum_{j\ne k} B_{jk} \hat a_j^\dag
  \hat a_k ,
\end{gather}
where $B_{jk} = \tfrac14 \gamma^2 (3\cos^2\theta_{jk}-1) / r_{jk}^3$,
$\gamma$ is the gyromagnetic factor, $\mathbf r_{jk} = \mathbf r_j -
\mathbf r_k$ is the distance between two nuclei, $\theta_{jk}$ is the
zenith angle of the vector $\mathbf r_{jk}$, and we used the first
order Holstein-Primakoff approximation.  At full polarization, we can rewrite the dipolar
Hamiltonian \eqref{eq:dipole} in terms of the bosonic spin wave mode
operators \eqref{eq:spin-modes} as $\hat H_D = \sum_{pq} \tilde B_{pq}
\hat\Phi_p^\dag \hat\Phi_q$, with $\tilde B_{pq} = \sum_{j\ne k}
B_{jk} \eta_{pj} \eta_{qk}^*$.  Thus, the storage mode is coupled to a
bath of non-storage modes as if it were a central spin coupled to a
mesoscopic spin bath \cite{prl88e186802&prb67e195329, prb74e035322}.
Although the total number of excitations is conserved, $\hat H_D$ is
responsible for decoherence of the qubit via transitions from the
storage state $\ket{\mathbf1}$ to non-storage states $\ket{1_q}$.  In
fact, the non-storage modes produce a fluctuating effective
transversal magnetic field with (complex) Larmor frequency $\hat
\Omega _{D-}^{\text{eff}} = 2 \sum_{q\ne0} \tilde B_{0q} \hat \Phi_q$.
If the electron were not present, these fluctuations would lead to a
decoherence rate $\Gamma_D \sim \Delta \Omega_D^{\text{eff}} = \big(
2\sum_{q\ne0} |\tilde B_{0q}|^2 \big) {}^{1/2}$ in the fully polarized
state, which is numerically found to be in the order of 100\,Hz for
GaAs (Fig.~\ref{fig:params}a).  With the protective gap, however, the
storage mode creation and annihilation operators ($\hat\Phi_0^\dag$
and $\hat\Phi_0$) rotate rapidly with respect to the other ones, and the
above coupling averages out and disappears in first order of the
dipolar perturbation.  In second order, the strength of the remaining
coupling between the storage mode and mode~$q$ is proportional to
$\Delta_\textgap^{-1} \sum_{r\ne0} \tilde B_{0r} \tilde B_{rq}$, and
the corresponding fluctuations yield a decoherence rate of $\Gamma_D'
\sim \Delta_\textgap^{-1} \big( 2\sum_{q\ne0} \big| \sum_{r\ne0}
\tilde B_{0r} \tilde B_{rq} \big|^2 \big) {}^{1/2} \sim
3 \times 10^4 \mathrm{Hz^2} / \Delta_\textgap$.  Typically,
$\Delta_\textgap \sim 1\,\mathrm{MHz}$ depending on the dot size
(Fig.~\ref{fig:params}a), so the effects of spin diffusion can be
suppressed by several orders of magnitude.

\paragraph{Non-perfect spin polarization.}%
Finally, we investigate the consequences of non-perfect nuclear spin
polarization.  It has been shown that partially polarized nuclei (at
finite temperature) can also be used for storing a qubit
state~\citep{prl91e246802}.  Instead of the fully polarized state
\eqref{eq:ket0}, the initial preparation drives the nuclear ensemble
into a statistical mixture of dark states $\ket{\mathcal D_{n,\beta}}$
defined by $\hat A_- \ket{\mathcal D_{n,\beta}} = 0$.  These dark
states can be characterized by the total number of spins flipped $n$
and the permutation group quantum number $\beta$.  As the detuning
$\delta$ is adiabatically swept from far negative to far positive, a
superposition of the $\ket\downarrow_e$ and $\ket\uparrow_e$ electron
spin states is mapped into the mixture of superpositions of the
nuclear spin states $\ket{\mathcal D_{n,\beta}}$ and
$\ket{\mathcal{E}_{n,\beta}} \equiv \frac{\mathcal{A}}{\Omega_n} \hat
A_+ \ket{\mathcal D_{n,\beta}}$, and the qubit state is efficiently
written into the memory~\citep{prl91e246802}.

When the electron is left in the quantum dot, it feels different
Overhauser fields for different dark states, hence the detuning should
be adjusted such that $\overline{\delta^{\text{OH}}_n +
  \delta^{\text{el}}} \gg \mathrm{Var} (\delta^{\text{OH}}_n)$.
Moreover, the hyperfine Rabi frequency also varies with $n$ and the
energy gap $\Delta_{\textgap,n}$ is not the same for all dark
states.  This inhomogeneous broadening would result in dephasing of
the qubit, but can be avoided by a symmetric spin echo sequence~\citep
{prl91e246802}.

To describe inhomogeneous effects in the case of non-perfect
polarization, first we note that the storage state
$\ket{\mathcal{D}_{n,\beta}}$ is no longer an eigenstate of the Knight
shift operator, but it is partially mapped into an orthogonal state:
$\hat H_K \ket{\mathcal{D}_{n,\beta}} = -\frac12 \delta^{\text{OH}}_n
\ket{\mathcal{D}_{n,\beta}} + \omega_n
\ket{\mathcal{D}_{n,\beta}^\perp}$.  This is due to the fact that the
inhomogeneous $\hat A_{z,\pm}$ operators do not follow the angular
momentum commutation relation.  Furthermore,
$\ket{\mathcal{E}_{n,\beta}}$ is neither an eigenstate of $\hat
H_\textgap$ nor of $\hat H_\textknight$: $\hat H
\ket{\mathcal{E}_{n,\beta}} = (-\frac12 \delta^{\text{OH}}_n +
\Delta_{K,n} + \Delta_{\textgap,n}) \ket{\mathcal{E}_{n,\beta}} +
\zeta_n \ket{\mathcal{E}_{n,\beta}^\perp}$.  The parameters can be
expressed as expectation values in~$\ket{\mathcal D_{n,\beta}}$:
\begin{gather}
  \Omega_n^2  = \mathcal A^2 
  \langle \hat A_- \hat A_+ \rangle, 
  \quad
  \omega_n^2 = \tfrac{\mathcal A^2}4 \big(
  \langle \hat A_z^2 \rangle - \langle \hat A_z \rangle^2 \big), 
  \nonumber\\
  \Delta_{\textgap,n} = {\mathcal A^4}
  \langle \hat A_- \hat A_+ \hat A_- \hat A_+ \rangle
  \big/ {4\delta_n \Omega_n^2},
  \nonumber\\
  \Delta_{\textknight,n} = \tfrac{\mathcal A}2 \langle\hat A_z\rangle
  - \mathcal A^3 \langle \hat A_- \hat A_z \hat A_+ \rangle
  \big/ {2 \Omega_n^2},
  \nonumber\\
  \zeta^2 = \bra{\mathcal E_{n,\beta}} 
  \hat H^2 \ket{\mathcal E_{n,\beta}} 
  - \bra{\mathcal E_{n,\beta}} 
  \hat H \ket{\mathcal E_{n,\beta}}^2 .
\end{gather}
The explicit form of the inhomogeneous dark
states~\citep{prl91e246802} allows us to estimate these values (see
Fig.~\ref{fig:params}b).  We expect that the storage mode is still
protected as long as $\omega_n$ and $\zeta_n$ are much smaller than
$\Delta_\textgap$, which is the case even for considerable unpolarized
fraction ($n/N$).

In summary, we have demonstrated that it is possible to suppress the
influence of spin-dephasing and spin-flips on a quantum memory
consisting of a delocalized ensemble of nuclear spins in a quantum dot
if the noise has a highly local character and the spectral width or
cut-off frequency of the noise spectrum is small compared to the
energy gap.  We have shown in particular that the memory can be
protected against nuclear spin diffusion mediated by dipole-dipole
interaction.  We have also analyzed the effects of inhomogeneous
hyperfine couplings and imperfect initial nuclear spin polarization.

JMT wishes to thank the Fleischhauer group for their kind hospitality
during his stay.  This work was supported by the EU network EMALI; JMT
is supported by Pappalardo.

\bibliographystyle{apsrev}
\bibliography{dotmem-refs}
\end{document}